\documentstyle[twoside,fleqn,espcrc2,graphicx,epsf]{article}
\newcommand{\beq}{\begin{equation}}
\newcommand{\eeq}{\end{equation}}

\title{Looking for Effects of Topology in the Dirac Spectrum of 
Staggered Fermions}

\author{
P.H. Damgaard\address{The Niels Bohr Institute/NORDITA, Blegdamsvej 17, 
DK-2100 Copenhagen, Denmark}, U.M. Heller\address{SCRI, Florida State 
University, Tallahassee, FL 32306-4130, USA}, R. Niclasen$^a$ and 
K. Rummukainen$^{a,}$\address{Helsinki Institute of Physics, P.O.Box 9, 00014 
University of Helsinki, Finland}
}
\begin{document}
\begin{abstract} 
We classify SU(3) gauge field configurations in different topological sectors
by the smearing technique. In each sector we compute the distribution
of low lying eigenvalues of the staggered Dirac operator.
In all sectors we find perfect
agreement with the predictions for the sector of topological charge zero.
The smallest Dirac operator eigenvalues of
staggered fermions at presently realistic lattice couplings are thus
insensitive to gauge field topology. On the smeared configurations, 
$4\nu$ eigenvalues go to zero in agreement with the index theorem. 
\end{abstract}
\maketitle
\section{Introduction}

In the finite-volume scaling regime $L \to \infty$ with $L << 1/m_{\pi}$
there are detailed analytical predictions for the rescaled microscopic
Dirac operator spectrum in gauge field sectors of fixed topological
charge $\nu$ \cite{SV,D,OTV}. We confront these predictions with quenched 
si\-mu\-la\-tions on staggered fermions, with lattice size $8^4$ and 
$\beta=5.1$. More details can be found in our paper \cite{DHNR}. 
 
Random Matrix Theory, or equivalently finite-volume partition
functions, can be used to compute \emph{exactly} the microscopic Dirac
operator spectral density
\beq
\rho_{s}(\zeta)=\lim_{V \rightarrow \infty} \frac{1}{V} \rho(\frac{\zeta}
{V\Sigma})
\eeq
where
\beq
\Sigma=\lim_{m \rightarrow 0} \lim_{V \rightarrow \infty}<\bar{\psi}\psi>=
\pi\rho(0)
\eeq
is the infinite-volume chiral condensate.

At finite lattice spacing, staggered fundamental fermions of SU(3)
gauge theory lead to a microscopic Dirac spectrum in the universality
class known as the chiral unitary ensemble (chUE).
In a quenched gauge field sector of topological charge $\nu$, this
microscopic spectral density reads
\beq
\rho_{s}^{(\nu)}(\zeta)=\pi \rho(0) \frac{\zeta}{2} 
\{J_{\nu}(\zeta)^{2}-J_{\nu-1}(\zeta)J_{\nu+1}(\zeta)\}
\eeq
As first observed by Verbaarschot \cite{V}, staggered fermions give good 
agreement with
analytical prediction of just the $\nu$=0 sector, even when
\emph{all} gauge field configurations are summed over.
To test whether staggered fermions at similar $\beta$-values
\emph{are} sensitive to topology at all, we have divided a large number of
configurations ($\sim$17,000) into different topological
sectors based on a variant of APE-smearing (see ref. \cite{DHNR} for 
the details).

\section{Classification scheme \& Analysis}

Using the naively latticized topological charge
\beq
\nu ~=~ \frac{1}{32\pi^2}\int\! d^4x {\mbox{\rm Tr}}[F_{\mu\nu}F_{\rho\sigma}]
\epsilon_{\mu\nu\rho\sigma} 
\eeq
a given gauge field configuration is assigned an integer topological
charge, $\nu$, if the rounded-off value of the `naive' $\nu$ is stable
between 200 and 300 APE-smearing steps. Other configurations
($\sim37\%$) are simply rejected.

\begin{figure}[ht]
\includegraphics[scale=0.4]{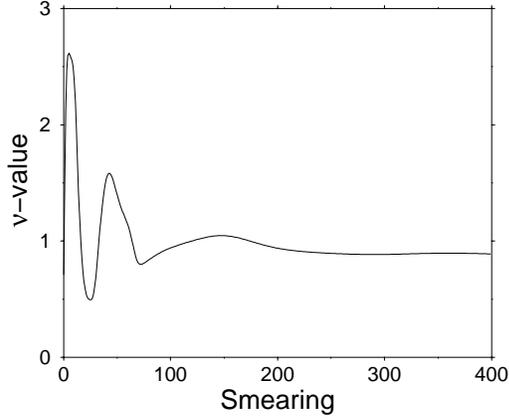}
\caption{An example of the smearing evolution
of the naive $\nu$-value.}
\label{fig1}
\end{figure}

\begin{figure}[!htb]
\includegraphics[scale=0.4]{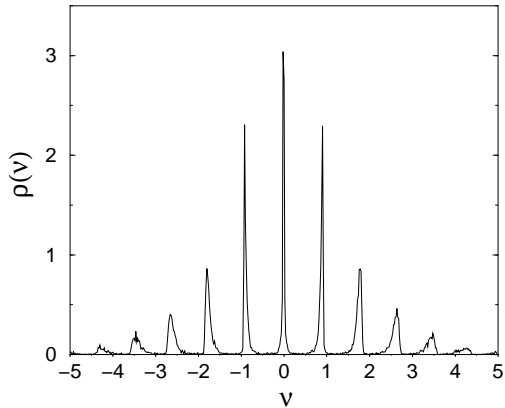}
\caption{The distribution of the naively measured
topological charge after 200 smearing steps.}
\label{fig2}
\end{figure}

The rounding-off of the naive $\nu$-values makes good sense, as can be seen
in figure \ref{fig2}. It shows the distribution of the measured
topological charge after 200 smearing steps. The distribution is
strongly peaked around quantized values, and our measured $\nu$-values
are rounded off to the obvious integer assignment.  
We see some
``renormalization'' of the topological charge as measured in this way,
but this is of no concern for us here. We shall only use the
measured $\nu$-values to make a classification of the {\em original},
un-smeared, configurations.

\begin{figure}[ht]
\includegraphics[scale=0.4]{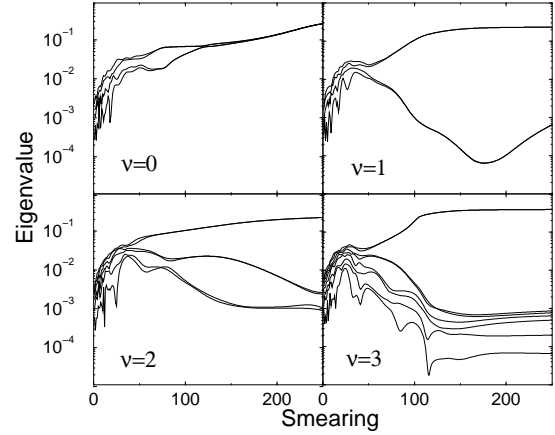}
\caption{The smearing history of the first few eigenvalues.}
\label{fig3}
\end{figure}

The staggered Dirac operator spectrum has a $\pm$ symmetry, and when we
trace a few small eigenvalues as a function of smearing in figure 
\ref{fig3}, we show only the positive ones.

As expected, we find that $4\cdot\nu$ eigenvalues become small
compared with the rest after very many smearing steps.  These are the
$\nu$ ``would be'' zero modes of 4 (continuum) flavors on the very
smooth configurations.

\begin{figure}[ht]
\includegraphics[scale=0.4]{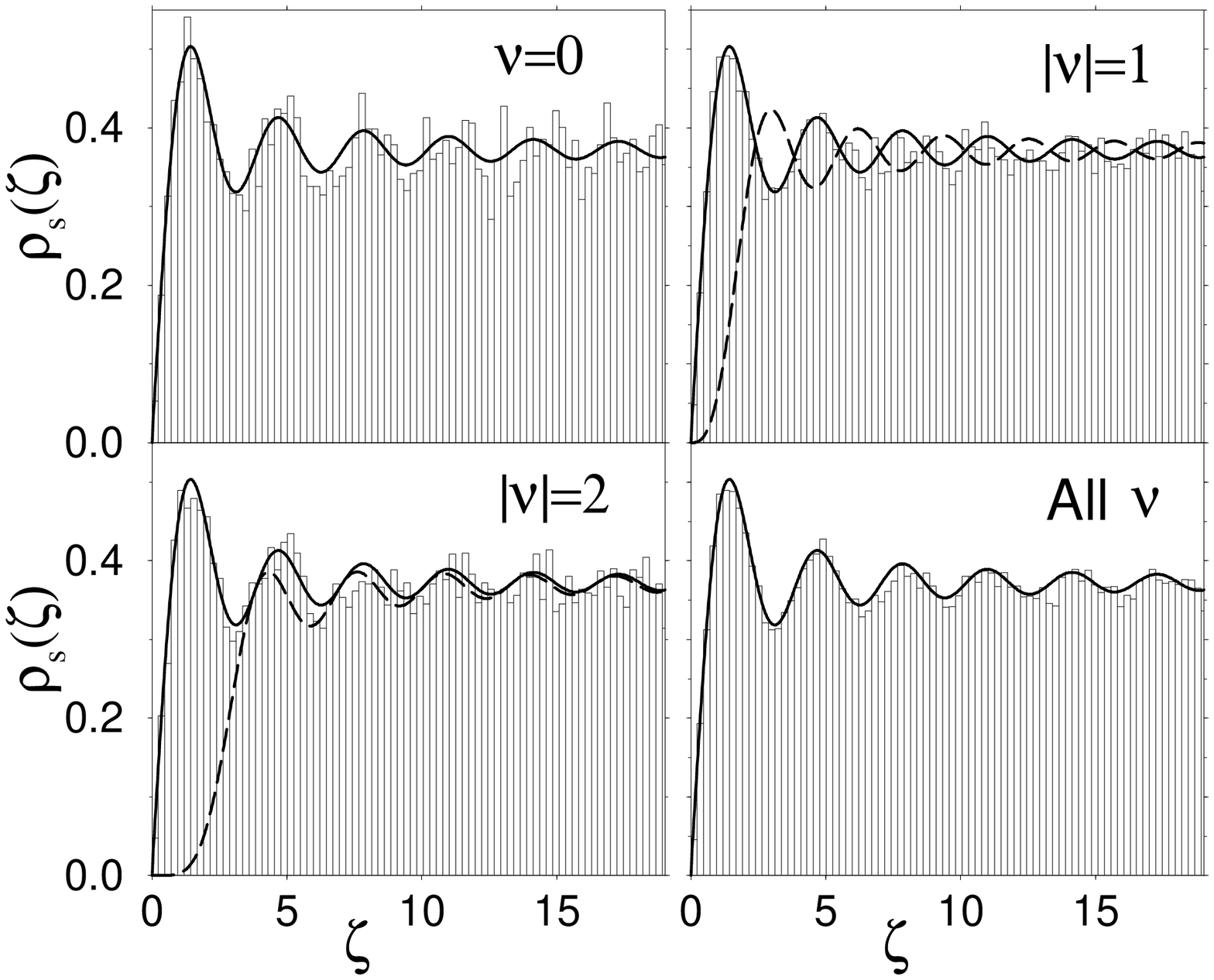}
\caption{Spectral densities configurations classified by $\nu$}
\label{fig4}
\end{figure}

All measurements of the microscopic Dirac ope\-ra\-tor spectrum are
performed on the original gauge field
configurations, classified into different
$\nu$-sectors according to the smeared values of $\nu$.
Shown in figure \ref{fig4} are the spectral densities obtained on
configurations classified by $\nu$ in that way.

Dashed lines are the analytical predictions for $\nu$=1 and 2 in
the appropriate graphs, and the solid curve is the analytical
prediction for $\nu$=0. At this $\beta$-value($\beta=5.1$) there is no
discernible deviation from the $\nu$=0 prediction even on
configurations that have been classified as $\nu=\pm 1,\pm 2.$

We have also compared the exact
predictions
\begin{eqnarray}
P_{min}^{\nu=0}(\zeta) & = &
\pi\rho(0)\frac{\zeta}{2}e^{-\frac{\zeta^2}{4}} \nonumber \\
\nonumber \\
P_{min}^{\nu=1}(\zeta) & = &
\pi\rho(0)\frac{\zeta}{2}I_2(\zeta)e^{-\frac{\zeta^2}{4}} \nonumber \\
\nonumber \\
P_{min}^{\nu=2}(\zeta) & = &
\pi\rho(0)\frac{\zeta}{2}\{I_2^2(\zeta)-I_1(\zeta)I_3(\zeta)\}e^{-\frac{\zeta^2}{4}}
\nonumber
\end{eqnarray}
for the distribution of just the smallest eigenvalue in the different
topological sectors. Even here we see no
deviation at all from the $\nu$=0 prediction.

It has recently been shown that it is possible to
recover the correct sensitivity of staggered fermions to topology
at very weak gauge coupling
(in the Schwinger model) \cite{Graz}. With fermions sensitive to gauge field
topology, nice agreement with the analytical predictions has been seen even 
far from the continuum limit, as it should be \cite{EHKN}.

\begin{figure}[ht]
\includegraphics[scale=0.4]{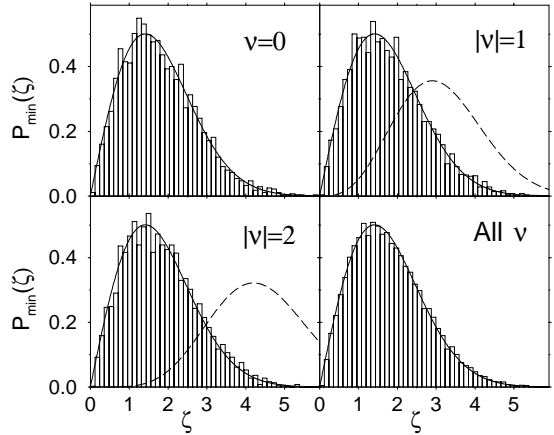}
\caption{Distributions of the smallest eigenvalue.} 
\label{fig5}
\end{figure}

The work of P.H.D. and K.R.
was partially supported by EU TMR grant ERBFMRXCT97-0122 and the work
of U.M.H by DOE contracts DE-FG05-85ER250000 and DE-FG05-96ER40979.
In addition P.H.D.
and U.M.H. acknowledge support by NATO Science Collaborative Research
Grant CRG 971487.
\vskip -3mm

\end{document}